



\def   \ni {\noindent}
\def   \cl {\centerline}
\def   \hmp {h^{-1}Mpc}

\def   \ssk {\vskip  5truept}

\def   \bsk {\vskip 15truept}

\def   \newline {\hfil\break}


\magnification=1000
\hsize 5truein
\vsize 8truein
\font\abstract=cmr8

\font\text=cmr10
\font\affiliation=cmssi10
\font\author=cmss10

\font\title=cmssbx10 scaled\magstep2

\def\ref{\par\noindent\hangindent 15pt}
\nopagenumbers
\null
\vskip 3.0truecm
\baselineskip = 12pt

{\title  THE DEBATE OF GALAXY CORRELATIONS
AND ITS THEORETICAL IMPLICATIONS
\ni
}


\bsk \bsk
{\author L. PIETRONERO $^{1}$, M MONTUORI$^{1,2}$ and
F. SYLOS LABINI$^{1,3}$
\ni
}

\bsk
{\affiliation                   
(1) Dipartimento di Fisica, Universita di Roma "La Sapienza", 00185 Roma, Italy
(2) Dipartimento di Fisica, Universita della Calabria, Italy
(3) Dipartimento di Fisica, Universita di Bologna, Italy
}

\bsk
\cl {\it (Received August 1995)}
\bsk
\baselineskip = 9pt

{\abstract
We discuss the problem of galaxy correlations by considering the various
 methods by
 which this information can be obtained. We focus in particular on
 the volume limited
 three dimensional samples and discuss a new way to increase the
scale of their statistical
validity. From our previous results and the most recent ones on all
the available
 catalogues we conclude that galaxy correlations show fractal
properties with dimension
$ D \approx  2$ up to the present observational limits ($200-800 \hmp$
depending
on the catalogue)
 without any tendency towards homogenization. We also discuss the
compatibility of
this result with the galaxy counts as a function of magnitude and the
angular catalogues
. A new picture emerges which changes the standard ideas about the
properties of the
 universe and requires a corresponding change in the theoretical
concepts that one should
 use to describe it.

\ni

}

\bsk

\baselineskip = 12pt

{\text

\ni 1. INTRODUCTION AND HISTORICAL PERSPECTIVE
\ssk

The first galaxy catalogues were only angular, namely they defined
the two angles
 corresponding to the galaxy positions in the sky (1). These angular
distributions show
 structures at small scales but appear rather smooth at large angular
scales. This situation
was therefore fully satisfactory with respect to the theoretical
expectations of a
homogeneous universe (2).
In the late seventies however the first redshift measurements
became available and
permitted the identification of the absolute distance of galaxies. In
this way it was
possible to obtain the complete three dimensional distribution of
galaxies. These
 distributions showed a more irregular structure with respect to the
angular data with the
appearance of superclusters and large voids. At first these large
structures were
considered as accidental or due to experimental incompleteness. But
more and more data
showed that the structures are all over and the voids do not fill with
better measurement.
This three dimensional picture did not show any more a clear
tendency towards homogenization and was in contrast with the angular data.
This conflictual situation found an apparent solution with the first
statistical analysis of
the CfA1 galaxy catalogue (3). In fact this analysis identified a small
correlation length of
only $\approx 5 \hmp$ in a catalogue that showed structures of much larger
sizes. The statistical
analysis appeared therefore to provide a way out such that large
structures can be
compatible with small correlation lengths.

In the following years the situation evolved in a dramatic way
because deeper and deeper
surveys showed larger and larger structures that appeared difficult to
reconcile with such a
small correlation length. In addition the catalogues of clusters gave a
correlation length of
$\approx 25 \hmp$, five times larger than the one of galaxies, even though the
clusters are made
themselves of galaxies.
This situation led to a great confusion and different authors looked
for different possible
solutions to the problem. At this stage various hypotheses were
formulated like the
luminosity segregation, the clustering richness relation that leads to
the biased theory of
structure formation etc.
In the end the most popular picture is that clusters and galaxy
structures require different
theories because their correlation show different amplitudes. A
linear theory is appropriate
for clusters while a non linear theory should be adopted for galaxies.
The large scale structures
can be compatible with small correlation lengths and with a large
scale homogeneity
because the amplitudes of the structures becomes smaller the
larger is the structure.
Finally a clear evidence of homogeneity cannot yet be obtained
because the present
samples are not yet fair.

In the past years we have challenged this picture (4) by showing
that it arises just by a
mathematical inconsistency in the characterization of the galaxy and
cluster correlations.
Our main result is that a correct analysis of the data shows fractal
correlations up to the
present observational limits. The galaxy-cluster mismatch disappears
and the visible
universe is characterized by a multifractal distribution of matter
when the galaxy masses
are also included. This requires a radical change of perspective for
the properties of the
universe and for the theoretical methods that one should use to
describe it. In this lecture
we present a colloquial discussion of these subjects including the
most recent results (5-
8).

\bsk
\ni 2. GALAXY CORRELATIONS
\ssk
The information about galaxy correlation is supposed to arise from a
variety of facts:

- The Cosmological Principle\
- The isotropy of the background radiation\
- The properties of angular catalogues\
- The galaxy number counts\
- The N-body simulations\
- The three dimensional galaxy catalogues\

The information obtained by these different sources is often
conflictual so it is important
to establish a hierarchy of validity and strength between the
different points mentioned
above. For example, suppose that the three dimensional distribution
of galaxies turns out
to be not homogeneous, what shall we do? Throw away the
telescopes or try to change
the Cosmological Principle?
And what about the background radiation? Should we disregard the
inhomogeneous
galaxies or the isotropic radiation?
In order to clarify the situation
it is useful to distinguish between conceptual
and technical
questions.
\bsk

{\it A. Conceptual Questions:}
\ssk

The {\it Cosmological Principle} corresponds to a reasonable
requirement of democracy:
we should try to avoid models that imply that we are in a very
special point in the
universe. Its interpretation in terms of isotropy and homogeneity for
the whole space is
however too strong. For example a fractal structure is locally
isotropic, in the sense that
all points (galaxies) have about the same environment, but it is not
homogeneous (4,9).
This occurs because isotropy implies homogeneity only for a regular
(analytical)
structure. A fractal is non-analytical everywhere so this relation does
not hold and we can
have a coexistence of democracy with non homogeneity. The
asymmetry in such a
structure is between occupied and empty points, but this is a
perfectly acceptable
asymmetry.

The {\it background radiation} and the observation of galaxy
positions are two different
experimental facts. If these two observations appear as conflictual
this means that we do
not have the correct theory to relate one to the other and not that we
should eliminate one
of the two or manipulate the data so that the disagreement is
eliminated. In this respect it
may be useful to try to separate clearly the bare experimental facts
from the theoretical
results. Generic statements of "consistency" in which the theory is
used directly in the
data analysis have usually led to confusion.

The {\it N-body simulations} correspond to computer experiments in
which the result is
crucially dependent on the assumptions for the meaning of the initial
and final elements,
the starting situation and the dynamical evolution. In order to
appreciate the subtleties
and possible complications of this type of simulations it is useful to
mention the present
state of the art in fully developed turbulence. Most authors believe
that this phenomenon
is certainly contained into Navier-Stokes equations, so in some sense
one knows the
correct theory. However even the most powerful simulations that
have been performed
with these equations do not seem to properly reproduce the
phenomenon. This is because
many different length scales are involved in the energy transfer and
dissipation and even
the most powerful computers do not seem to be able to fully describe
this phenomenon.
In this respect one should be extremely careful in the interpretation
of these simulations
with respect to the physical reality.
\bsk

{\it B. Technical Questions}
\ssk

Clearly the most direct information about galaxy correlations comes
from {\it three
dimensional volume-limited samples}. The limited amount of data
may pose a problem
of statistical validity. However, within the limits of their statistical
validity, these samples
represent directly the real properties of galaxy correlations while all
the other
measurements can only lead to consistency arguments. In case of
disagreement the $3-d$
catalogues should be used as a test for the other measurements.
The $3-d$
 volume limited catalogues have usually been analyzed in
terms of the so called
two point correlation function $\xi(r)$ (10). The characteristic correlation
length is defined by
the distance at which this function equals unity ($\xi(r_0)=1$).
 In the past we have
discussed in detail
the inconsistency of this approach (4) and the fact that the
characteristic lengths obtained
represented just a fraction of the sample volume and not the real
correlation length. A
new analysis of the data with more general methods leads in fact to the
result that galaxy and
cluster catalogues show fractal correlations up to their limits without
any evidence for
homogenization (4). The statistical limits of these analyses
correspond roughly to the
radius of the largest sphere that can be contained in the sample. For
galaxies this is about
$ 20-30 \hmp$ and for clusters about $ 80 \hmp$. These studies already
allowed us to show that
the claimed correlation lengths of 5 and 25 $\hmp$ for galaxies and
clusters respectively
were spurious results due to the inconsistent analysis. This situation
eliminates the
apparent inconsistency between galaxy and cluster correlations and
made the correlation
analysis compatible with the observation of large scale structures.

Some authors tried to push the statistical limits of these catalogues
beyond the above
limits by using various types of weighting schemes (11). In our
opinion this is a very
risky procedure because, in one way or another, all the weighting
schemes imply some
homogeneity hypothesis. If this would not be so, then from the
observation of a few
galaxies, one would be able to reconstruct the properties of the entire
universe.

In order to push the limiting length of statistical validity for the 3-d
samples we have
followed instead a different path that again does not involve any a
priori assumption. The
basic idea is that if one has a large volume limited sample and one
integrates from the
observation point the galaxies at progressively larger distances one
should observe a
power law with exponent 3 for the homogeneous case or a lower
exponent for a fractal
distribution. Clearly this is what is done from each galaxy in the
complete correlation
studies. However in that case one uses spherical shells for the
integration and from this
comes the previous limitation. The integration from the origin uses
instead the conic sample that
can be properly defined only from this point. In this way the
statistical fluctuations,
especially at small scale, are very large but the advantage is that, in
the favourable
situations, the effective depth can be extended by more than a factor
of four.

We have performed this type of analysis in the Perseus-Pisces
survey, in CfA1 (just as a
test) and in the ESP survey. We report below a comprehensive
summary of our
correlation studies of both types for the various catalogues:

- CfA1 and CfA2. For CfA1 the analysis are limited to
20 $\hmp$ and show
fractal properties up to this limit. This trend continues in CfA2 in
view of the shift of $r_0$
with depth and on the properties of its power spectrum (4,7).
Therefore both CfA1 and 2
show fractal correlations up to their limits. The fractal dimension is
somewhat smaller at
small scales and it becomes about 2 for CfA2.

- Perseus-Pisces. For the correlation function the limit is 30$\hmp$
while for the counting
from the origin one can extend this limit up to 130$\hmp$. Both data are
consistent with a
fractal distribution with dimension $D = 2$. (6).

- LEDA database. Correlation analysis up to 150 $\hmp$. Fractal with $D =
2$. Scaling of $r_0$
clearly defined up to 60$\hmp$  (8).

- ESP survey (12)(preliminary results). Correlation analysis
is impossible because of the too
small solid angle. Integral from the vertex shows fractal properties
up to 800 $\hmp$(5).

A summary of the behavior of the conditional density as a function
of distance for the
various catalogues is shown in Fig.1, with the exception of the
ESP data that are still preliminary.

\midinsert
\vskip 0.5truecm
\par\noindent

\leftskip=1truecm
\rightskip=1truecm
\noindent
{\abstract
FIGURE 1.
$\Gamma(r)$ for various catalogues. The amplitude in the various
cases is {\it not} arbitrary, and it is normalized
only to take into account the different luminosity
distribution in different volume limited samples (6),(8)}
\vskip 0.5truecm
\leftskip=0truecm
\rightskip=0truecm
\endinsert

The result is that in all cases galaxy correlation are well defined and
extend up to the
observational limits without any tendency towards homogenization.
As a byproduct of
these studies we can also make the following comments:

- The so called luminosity segregation hypothesis can be definitely
eliminated. The shifts
of $r_0$ are clearly due to the sample depth as predicted for a fractal
distribution.

- The use of weighting schemes should be avoided because it is now
clear that they were
responsible for the apparent trends towards homogenization (11)
that have been
disproved by the new results of galaxy counts for volume limited
samples.

Another way to gain information about the galaxy distribution
 without knowing the
redshifts is the {\it number count as a function of apparent
magnitude} and the
corresponding {\it angular catalogues}. In these cases however the
information is not
direct and it requires a nontrivial interpretation.
The galaxy count as a function of magnitude represents one of the
very first elements that
have been studied (13) and gave rise to extensive problems that are
still unclear
today(2,5,14). The point is that one can easily show that the
exponent which characterizes
this counting should be $\alpha = D/5$. This implies $\alpha = 0.6$
 for a
homogeneous distribution and
a smaller value for a fractal one. The observation show the very
puzzling behavior that $\alpha = 0.6$
 for relatively small scales while for larger scales one has
$\alpha = 0.4$
 up to the limits.
Eventually one
would have expected just the opposite situation, namely fractal
behavior at relatively
small scales followed by homogeneity at very large scales. At the
moment the traditional
interpretation of these data is that the value 0.6 corresponds to the
much hoped
homogeneity, while the smaller value at large scales is affected by
evolutionary effects
invoked for the occasion.

We now know for sure that this
interpretation cannot be correct.
In fact it is clear beyond any doubt (from the $3-d$
 volume limited
catalogues) that fractal
correlations extend up to $ \approx 100 \hmp$
 and probably much more. At these
 relatively small
scales evolutionary effects are certainly negligible. So we have the
situation that a fractal
distribution with dimension $D = 2$ leads to a counting vs. magnitude
 behavior that appears
instead to correspond to $D = 3$.
We have studied this effect in detail
by considering both
real catalogues and computer simulations with preassigned
properties and our conclusion
is that these counting are strongly affected by finite size effects
(15). Namely at
relatively small scales one finds almost no galaxies because the total
number is rather
small. Then one enters in a regime dominated by finite size
fluctuation effects. Finally the
correct scaling behavior of the distribution is recovered. This means
for example that if
one has a fractal distribution, there will be first a raise of
the conditional density,
due to finite size effects
because no galaxies are present before a certain characteristic length.
Once one enters in
the correct scaling regime the
density will begin to decay with the correct
power law as corresponding to the
fractal correlations.
 So in this intermediate regime of raise and fall the will
be a region in which
the density can be roughly approximated by a constant value.
This region will lead to
an apparent dimension $D = 3$,
which however is not real but just due
to the finite
size effects. We have checked that this is exactly what happens in the
real counting in
which the initial exponent $\alpha = 0.6$ corresponds to the finite size
effects and the subsequent
value $\alpha = 0.4$ is the real value, now perfectly consistent with the
observed fractal
dimension $D = 2$.

This counting problem is present as well in the calculation of the
amplitude of the angular
correlation function. This quantity also correspond to the integration
from a single point
and therefore is also strongly affected by these finite size
fluctuations. Therefore the
scaling argument at the basis of the presumed homogeneity inferred
from the angular
catalogues corresponds to these finite size effects and not to the real
scaling properties of
the distribution.
We hope this discussion clarifies now completely the statistical
properties of galaxy and
clusters and will permit to focus on the real theoretical problems.

\bsk

\ni 3. AMPLITUDES VERSUS EXPONENTS
\ssk

{}From what we have seen all we can say about the galaxy distribution
is that it is a fractal
up to the present observational limits. This means that it is not
possible to define concepts
like the average density of galaxies. Therefore also the amplitude of
the correlation
function has no physical meaning because one cannot define what is
small and what is
big. In the same way it is not possible to assign any meaning to the
relative density
fluctuation $\delta N/N$ because the normalization value N is not
intrinsic. Clearly $\delta N/N$
goes to zero at the sample limits also for a fractal structure, as shown
in Fig.2

\midinsert
\vskip 0.5truecm
\par\noindent

\leftskip=1truecm
\rightskip=1truecm
\noindent
{\abstract FIGURE 2. Behavior of $\delta N/N$ as a function of size
 $r$ in a portion of a fractal structure for various depths of the sample
$R_s=100, 200, 300 \hmp$. The average density is computed
in the whole sample of radius $R_s$. The fact that  $\delta N/N$
tends to zero does not mean that the fluctuations
are small and an homogenous distribution has been reached. The
distance at which
$\delta N/N=1$ scales with sample depth and has not physical meaning.
}
\vskip 0.5truecm
\leftskip=0truecm
\rightskip=0truecm
\endinsert

but this does not mean that the system is becoming homogeneous.
This situation may appear strange but it is indeed quite common in
various fields of
physics where one deals with scale-invariant or self-similar
structures (4). In these cases
the relevant physical phenomenon that leads to the scale-invariant
structures is
characterized by the exponent and not by the amplitude.
Correspondingly one has to
change the theoretical framework into one that it capable of dealing
with non analytical
fluctuations. This means going from differential equations to
something like the
Renormalization Group for the study of the exponents.

If a crossover
towards
homogeneity would eventually be detected, this would not change
the above discussion
but it would simply introduce a crossover into it. The fractal nature
of the observed
structures would, in any case require this change of theoretical
perspective.
\bsk

\vskip 0.1truecm
\ni {REFERENCES}
\ssk

\ref 1. For example see: C.D. Shane and C.A. Wirtanen, Pub. Lick Obs. 22,
part 1, (1967); F.
 Zwicky et al., Cal.Inst. Tech., "1961-1968 Catalogues of Galaxies and
Clusters".
\ref 2. P.J.E. Peebles, "Principles of Physical Cosmology", Princeton Univ.
Press (1993).
\ref 3. M. Davis and P.J.E. Peebles, Ap. J. 267, 465 (1983)
\ref 4. P.H. Coleman and L. Pietronero, Physics Reports 213, 311 (1992)
\ref 5. Yu.V. Baryshev, F. Sylos Labini, M. Montuori and L. Pietronero,
Vistas in Astronomy
38, 419 (1994)
\ref 6. F. Sylos Labini, M. Montuori and L. Pietronero, this issue p. xxx
\ref 7. F. Sylos Labini and L. Amendola, this issue, p. xxx
\ref 8. H. Di Nella and F. Sylos Labini, this issue, p. xxx
\ref 9. F. Sylos Labini, Ap. J. 433, 464 (1994)
\ref 10. M. Davis and P.J.E. Peebles, Ap. J. 267, 465 (1983)
\ref 11. L. Guzzo et al, Ap. J. Lett. 382, L5 (1992)
\ref 12. E. Zucca et al. This issue, p. xxx
\ref 13. E. Hubble, Nat. Acad. Sci. Proc., 15, 168 (1929)
\ref 14. F. Sylos Labini, A. Gabrielli, M. Montuori and L. Pietronero,
preprint.
}
\end